\documentclass[10pt,letterpaper,twocolumn]{article} 

\usepackage{ol2}
\usepackage[draft]{hyperref}
\usepackage{amsmath}

\begin{document}

\twocolumn[ 

\title{Hyperfine Paschen-Back regime realized in Rb nanocell}


\author{A. Sargsyan,$^{1}$ G. Hakhumyan,$^{1,2}$ C. Leroy$^{2,*}$, Y. Pashayan-Leroy$^2$, A. Papoyan$^{1}$ and D. Sarkisyan$^{1}$}

\address{
$^1$Institute for Physical Research, NAS of Armenia, Ashtarak-2, 0203, Armenia\\

$^2$Laboratoire Interdisciplinaire Carnot de Bourgogne, UMR CNRS 6303, Universit\'{e} de Bourgogne, \\
21078 Dijon Cedex, France \\

$^*$Corresponding author: claude.leroy@u-bourgogne.fr
}

\begin{abstract}A simple and efficient scheme based on one-dimensional
nanometric thin cell filled with Rb and strong permanent ring magnets
allowed direct observation of hyperfine Paschen-Back regime on \textit{D}$_1$
line in $0.5 - 0.7$~T    magnetic field. Experimental results are
perfectly consistent with the theory. In particular, with  $\sigma^{+}$
laser excitation, the slopes of $B$-field dependence of frequency
shift for all the 10 individual transitions of $^{85,87}$Rb are the same
and equal to $18.6$~MHz/mT. Possible applications for magnetometry
with submicron spatial resolution and tunable atomic frequency
references are discussed.\end{abstract}


 ] 

\noindent
Rubidium atoms are widely used in atomic cooling, information storage,
spectroscopy, magnetometry etc \cite{Budker,Fleisch}. Miniaturization of alkali vapor
cells is important for many applications \cite{Knappe,Sarkisyan1,Urban,Leroy1}.
\begin{figure}[b]
\centerline{\includegraphics[width=8.3cm]{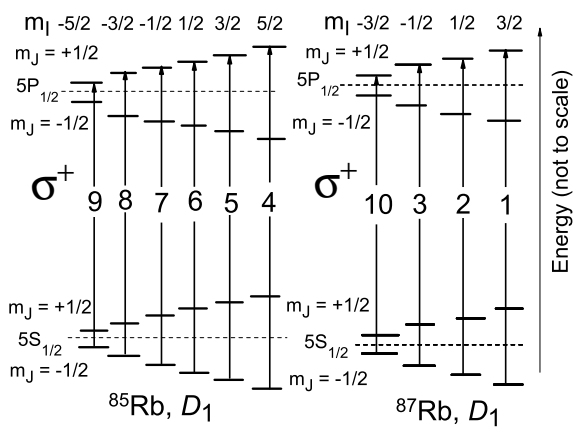}}
\caption{Diagram of $^{85}$Rb ($I = 5/2$) and $^{87}$Rb ($I = 3/2$) transitions
for  $\sigma^+$ laser excitation in HPB regime. The selection rules:  $m_J = +1$;
 $m_I = 0$.}
 \label{fig:Fig1}
\end{figure}
 Atom located in magnetic
field undergoes shift of the energy levels and change in transition probabilities,
therefore precise knowledge of the behavior of atomic transitions is very important~\cite{Tremblay}.
In case of alkali atomic vapor use a sub-Doppler resolution is needed to  study separately
each individual atomic transition between hyperfine (hf) Zeeman sub-levels of the ground
and excited states (in case of a natural mixture of $^{85}$Rb and $^{87}$Rb
the number of closely spaced atomic transitions can reach several tens).
Recently it was shown that a one-dimensional nanometric-thin cell (NTC) with the
thickness of Rb atomic vapor column $L = \lambda$, where  $\lambda= 794$~nm is the wavelength
of laser radiation resonant with \textit{D}$_1$ line of Rb, is a good tool to obtain sub-Doppler
spectral resolution. Spectrally narrow velocity-selective optical pumping (VSOP)
resonances located exactly at the position of atomic transitions appear in the
 transmission spectrum of NTC at laser intensities $\cong 10$~mW/cm$^2$~\cite{Sarkisyan1,Leroy1,Leroy2}.
When NTC
is placed in a weak magnetic field, the VSOPs are split into several components
depending on total angular momentum quantum numbers $F = I + J$, with amplitudes
and frequency positions depending on $B$-field, which makes it convenient to study
separately each individual atomic transition.
\begin{figure}[b]
\centerline{\includegraphics[width=8.3cm]{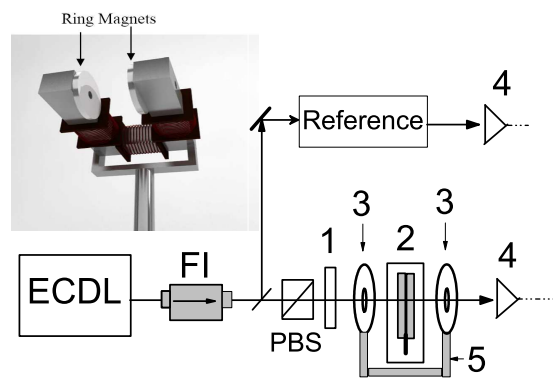}}
\caption{Sketch of the experimental setup. ECDL - diode laser, FI - Faraday isolator,
\textit{1} -  $\lambda/4$ plate, \textit{2} - NTC in the oven, PBS - polarizing beam splitter,
\textit{3} - permanent ring magnets,
\textit{4} - photodetectors, \textit{5} - stainless steel $\Pi$-shape holder (shown in the inset).}
\label{fig:Fig2}
\end{figure}
 \indent In this Letter we describe a simple and robust system
 based on NTC and permanent magnets, which allows of achieving magnetic
 field up to $0.7$~T sufficient to observe a hyperfine Paschen-Back regime~\cite{Alexandrov}.
 The magnetic field required to decouple the nuclear and electronic
 spins is given by $B \gg A_{hfs} /\mu_B$ $ \cong 0.2$~T for $^{87}$Rb, and
 $\cong 0.07$~T for $^{85}$Rb,
where $A_{hfs}$ is the ground-state hyperfine coupling coefficient for $^{87}$Rb
and $^{85}$Rb and $\mu_B$ is the Bohr magneton~\cite{Olsen}. For such a large magnetic
field the eigenstates of the Hamiltonian are described in uncoupled
basis of $J$ and $I$ projections ($m_J, m_I$). In Fig.~\ref{fig:Fig1} six transitions of $^{85}$Rb
labeled $\textit{4 - 9}$, and four transitions of $^{87}$Rb labeled $\textit{1 - 3}$ and $\textit{10}$ are shown
in the case of  $\sigma^+$ polarized laser excitation for the HPB regime.\\
\begin{figure}[htb]
\centerline{\includegraphics[width=8.3cm]{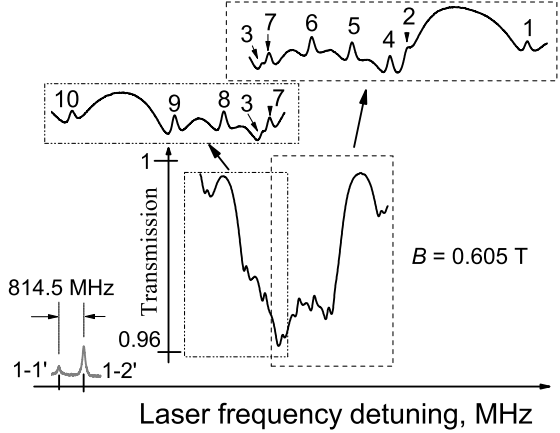}}
\caption{Transmission spectrum of Rb NTC with $L = \lambda$  for $B = 0.605$~T and
$\sigma^+$ laser
excitation. Well resolved VSOP resonances located at atomic transitions are
labeled \textit{1 - 10} (six transitions, \textit{4 - 9} belonging to $^{85}$Rb,
and four transitions, \textit{1,2,3} and \textit{10} belonging to $^{87}$Rb).
Change in probe transmission is
$\Delta T = 4\%$. The lower gray curve is the fluorescence spectrum of the reference NTC
with $L =  \lambda/2$, showing the positions of $^{87}$Rb, $F_g=1 \rightarrow F_e=1,2 $
transitions for $B = 0$, labeled as $\textit{1} - \textit{1}^\prime$ and $\textit{1} - \textit{2}^\prime$
(all frequency shifts are measured from $F_g=1 \rightarrow F_e=2$).}
\label{fig:Fig3}
\end{figure}
\indent The sketch of the experimental setup is shown in Fig.~\ref{fig:Fig2}. The circularly ($\sigma^+$)
polarized beam of extended cavity diode laser (ECDL,  $\lambda = 794$~nm, $P_L = 30$~mW,
$\gamma_L < 1$~MHz) resonant with Rb \textit{D}$_1$ line, was directed at normal incidence onto
the Rb NTC \textit{(2)} with the vapor column thickness $L = \lambda  = 794$~nm (a typical
example of a recent version of the NTC is described in~\cite{Leroy1}).
\begin{figure}[htb]
\centerline{\includegraphics[width=8.3cm]{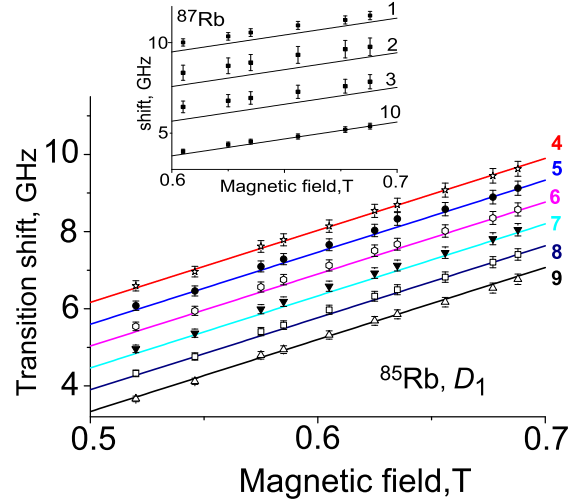}}
\caption{Magnetic field dependence of frequency shift for transition
components labeled \textit{4 - 9} ($^{85}$Rb) and \textit{1 - 3, 10} ($^{87}$Rb, in the inset).
Solid lines: theory; symbols: experiment. The inaccuracy is $\leq 2 \%$
for components \textit{4 - 9, 1} and \textit{10}, and $\leq 5 \%$ for components \textit{2} and \textit{3}.
Larger error for \textit{3} and \textit{2} is caused by closely located strong
transition \textit{7} and location on the transmission spectrum wing, respectively.}
\label{fig:Fig4}
\end{figure}
The NTC was placed in a special oven with two openings. The transmission signal
was detected by a photodiode \textit{(4)} and was recorded by a four-channel
digital storage oscilloscope. A polarizing beam splitter (PBS) was used to purify initial
linear radiation polarization of the laser radiation followed by
a $\lambda /4$ plate \textit{(1)} to produce a circular polarization. The magnetic
field was directed along the laser radiation propagation direction $\textbf{k}$.
About $30\%$ of the pump power was branched to the reference unit with an
auxiliary Rb NTC. The fluorescence spectrum of the latter at $L =  \lambda/2$ was
used as a frequency reference for $B = 0$. The magnetic field was measured by
a calibrated Hall gauge. Extremely small thickness of NTC is advantageous
for application of very strong magnetic fields by permanent ring magnets (PRM)
otherwise unusable because of strong inhomogeneity of magnetic field: in NTC,
the variation of $B$-field inside atomic vapor column is less by several orders
than the applied $B$ value. The ring magnets are mounted on a $50 \times 50$~mm$^2$ cross
section  $\Pi$-shaped holder made from soft stainless steel (see inset in Fig.~\ref{fig:Fig2}).
Additional form-wounded Cu coils allow of applying extra $B$-field (up to $\pm0.1$~T).
NTC is placed between PRMs. The linearity of the scanned frequency was
tested by simultaneously recorded transmission spectra of a Fabry-P\'{e}rot
etalon (not shown). The nonlinearity has been evaluated to be about $1\%$
throughout the spectral range. The imprecision in the measurement of the absolute
$B$-field value is   $\pm5$~mT. The recorded transmission spectrum of the Rb NTC with
$L = \lambda$  for  $\sigma^+$ laser excitation and $B = 0.605$~T is shown in Fig.~\ref{fig:Fig3}.
The VSOP resonances labeled $\textit{1 - 10}$ demonstrate increased transmission at the positions
of the individual Zeeman transitions. In the case of HPB the energy of the
ground $5S_{1/2}$ and upper $5P_{1/2}$ levels for Rb \textit{D}$_1$ line is given
by a simple formula~\cite{Alexandrov}:
\begin{equation}
\label{Eq1}
E_{|J,m_J,I,m_I\rangle}=A_{hfs}m_Jm_I+\mu_B(g_Jm_J+g_Im_I)B.
\end{equation}
The values for nuclear ($g_I$) and fine structure ($g_J$) Land\'{e} factors, and
hyperfine constants $A_{hfs}$ are given in~\cite{Steck}. The magnetic field dependence
of frequency shift for components \textit{4 - 9 }($^{85}$Rb) is shown in Fig.~\ref{fig:Fig4}
(solid lines: HPB theory; symbols: experiment, inaccuracy does not exceed $2\%$).
The similar dependence for $^{87}$Rb (components \textit{1 - 3} and \textit{10})
are shown in the inset of Fig.~\ref{fig:Fig4}. The HPB regime condition is fulfilled better for
$B > 0.6$~T. As it is seen from Eq.~(\ref{Eq1}), and also confirmed  experimentally,
the dependence slope is the same for all the transition components of both $^{85}$Rb
and $^{87}$Rb: [$g_J(5S_{1/2}) m_J + g_J (5P_{1/2}) m_J$]  $\mu_BB = 18.6$~MHz/mT
(as $g_I \ll g_J$  we ignore its contribution). Onset of this value is indicative
of Rb \textit{D}$_1$ line HPB regime. Note that in our previous study for
$B \sim 20$mT~\cite{Sarkisyan2}
and under the same conditions we observed 32 transitions, as opposed to 10
remaining in HPB regime.\\
\indent Rb NTC could be implemented for mapping strongly inhomogeneous
magnetic fields by local submicron spatial resolution. Particularly,
for $0.1$~T/mm gradient, the displacement of NTC by $5$~$\mu$m results in
$10$~MHz frequency shift of VSOP resonance, which is easy to detect.
Also development of a frequency reference based on NTC and PRM, which
is $B$-field-tunable in over $10$~GHz range, is of high interest.
The above studies and techniques can be successfully implemented also
for HPB studies of \textit{D} lines of Na, K, Cs, and other atoms.\\

\indent The research leading to these results has received funding from
the FP7/2007-2013 under \textit{grant agreement} n$^\circ$~205025 - IPERA. Research
conducted in the scope of the International Associated Laboratory IRMAS
(CNRS-France \& SCS-Armenia).



\begin{thebibliography}{99}


\bibitem{Budker} D. Budker, D. F. Kimball, and D. P. DeMille, \textit{Atomic Physics} (Oxford Univ. Press, Oxford, 2004).
\bibitem{Fleisch} M. Fleischhauer, A. Imamoglu, and J. P. Marangos, "Electromagnetically induced transparency: Optics in coherent media"
Rev. Mod. Phys. \textbf{77,} 633-–673 (2005).
\bibitem{Knappe}S. Knappe, L. Hollberg, J. Kitching, "Dark Line Resonances in Sub-Millimeter Structures" Opt. Lett. \textbf{29,} 388--390 (2004).
\bibitem{Sarkisyan1} 	A. Sargsyan, G. Hakhumyan, A. Papoyan, D. Sarkisyan, A. Atvars, M. Auzinsh,
"A novel approach to quantitative spectroscopy of atoms in a magnetic field and applications based on an atomic vapor cell with
$L = \lambda$"  Appl. Phys. Lett. \textbf{93,} 021119 (2008).
\bibitem{Urban} 	T. Baluktsian, C. Urban, T. Bublat, H. Giessen, R. L\"{o}w, T. Pfau,
"Fabrication method for microscopic
vapor cells for alkali atoms" Opt. Lett. \textbf{35,} 1950--1952 (2010).
\bibitem{Leroy1} 	G. Hakhumyan, C. Leroy, Y. Pashayan-Leroy, D. Sarkisyan, M. Auzinsh,
"High-spatial-resolution monitoring of strong magnetic field using Rb vapor nanometric-thin cell" Opt. Commun. \textbf{284,} 4007--4012 (2011).
\bibitem{Tremblay} P. Tremblay, A. Michaud, M. Levesque, S. Th\'{e}riault, M. Breton, J. Beaubien, and N. Cyr, "Absorption profiles of alkali-metal \textit{D} lines
in the presence of a static magnetic field" Phys. Rev. A \textbf{42,} 2767--2773 (1990).
\bibitem{Leroy2}G. Hakhumyan, A. Sargsyan, C. Leroy, Y. Pashayan-Leroy, A. Papoyan, D. Sarkisyan, "Essential features of optical processes in neon-buffered
submicron-thin Rb vapor cell" Opt. Express \textbf{18,} 14577--14585 (2010).
\bibitem{Alexandrov}	E. B. Alexandrov, M. P. Chaika, G. I. Khvostenko, \textit{Interference of Atomic States} (Springer-Verlag, Berlin, 1993).
\bibitem{Olsen} B. A. Olsen, B. Patton, Y.-Y. Jau, and W. Happer, "Optical pumping and spectroscopy of Cs vapor at high magnetic field"
Phys. Rev. A \textbf{84,} 063410 (2011).
\bibitem{Steck}D. A. Steck, "Rubidium 85 \textit{D} line data, Rubidium 87 \textit{D} line data", http://steck.us/alkalidata.
\bibitem{Sarkisyan2} D. Sarkisyan, A. Papoyan, T. Varzhapetyan, K. Blu\v{s}s, M. Auzinsh,
"Fluorescence of rubidium in a submicrometer vapor cell: spectral resolution of atomic
transitions between Zeeman sublevels in a moderate magnetic field" J. Opt. Soc. Am. B \textbf{22,} 88--95 (2005).
\end{thebibliography}
\end{document}